\documentstyle[twocolumn,psfig,aps]{revtex}
\begin{document}
\draft
\twocolumn[\hsize\textwidth\columnwidth\hsize\csname @twocolumnfalse\endcsname
%
%
%

\title{Diagonalization in Reduced Hilbert Spaces
using a Systematically Improved Basis: Application to Spin Dynamics in
Lightly Doped Ladders}

\author{E. Dagotto$^1$, G. B. Martins$^1$, J. Riera$^2$, A. L. 
Malvezzi$^1$, and C. Gazza$^1$}

\address{$^1$ National High Magnetic Field Lab and Department of Physics,
Florida State University, Tallahassee, FL 32306}
\address{$^2$ Instituto de F\'{\i}sica Rosario, Avenida Pellegrini 250, 
2000 Rosario, Argentina}

\maketitle

\begin{abstract}

A method is proposed to improve the accuracy of approximate techniques 
for strongly correlated electrons  that use reduced
Hilbert spaces. As a first step, the method involves a
change of basis that incorporates exactly part of the 
short distance interactions. 
The
Hamiltonian is rewritten in new variables that better represent
the physics of the problem under study. 
A Hilbert space expansion
performed in the new basis
follows. The method is successfully
tested using both the Heisenberg model
and the $t-J$ model with holes on 2-leg ladders and chains, 
including estimations for ground state energies, static correlations, and spectra of
excited states. 
An important feature of this technique is its
ability to calculate dynamical responses on clusters  
larger than those
that can be studied using Exact Diagonalization. The method
is applied to the analysis of the dynamical spin structure factor 
$S({\bf q},\omega)$ on clusters with $2 \times 16$ sites and 0 and 2
holes. Our results confirm previous studies (M. Troyer, H. Tsunetsugu,
and T. M. Rice, Phys. Rev. ${\bf B 53}$, 251 (1996)) which suggested that the
state of the lowest energy in the spin-1 2-holes subspace corresponds to
the bound state of a hole pair 
and a spin-triplet. Implications of this result for neutron scattering
experiments both on ladders and planes are discussed.
\end{abstract}
\vskip2pc]
\narrowtext

%
%

\section{Introduction}

The study of strongly correlated electrons is currently among the most active
areas of research in condensed matter physics. In recent years the
discovery of a variety of new phenomena
such as high temperature
superconductivity, buckyballs, and colossal magnetoresistance
effects~\cite{bednorz} have
triggered a huge theoretical effort devoted to the analysis of the
electronic 
models proposed for these compounds. 
Since  experimental data suggests that the electronic  
interactions cannot be considered as small perturbations
in the typical 
regime of couplings of these models,
the use of nonperturbative techniques, especially 
computational ones,
has become a popular approach for the study of 
complicated many-body problems in the area of materials
research.

In spite of their popularity, numerical methods are not without
problems. The well-known Lanczos 
approach~\cite{review}
provides exact static and dynamical information for finite clusters.
However, current memory limitations in available computers constrain
the cluster sizes that can be studied.
An alternative is
the same Lanczos method but now
applied on a reduced basis set. However, experience
has shown that the convergence to accurate results is slow, 
at least for the basis expansion procedures proposed thus 
far~\cite{wenzel,jose}. Recently,
the Density Matrix Renormalization Group (DMRG) 
has been introduced~\cite{white}
providing an optimal way to perform a basis expansion in
quasi-one dimensional systems. The method
certainly 
gives accurate results in the study of equal-time correlations
for a
variety of models, but currently it does not
provide dynamical information 
and since it is typically applied on open
boundary clusters the study of the momentum
dependence of observables is complicated. Alternative
techniques such as the Quantum Monte Carlo algorithm~\cite{qmc},
where a guided sampling of configurations is carried out,
can handle medium size clusters but it fails at low temperatures 
 due to the ``sign problem''. 

Some of the complications 
of these algorithms can be traced back to the
important differences between the degrees of freedom used explicitly
in the construction of
the Hamiltonians, and those that are dynamically generated in the
solution of the problem. A typical example 
is provided by the problem of holes close to
half-filling in the $t-J$ model: the eigenstates are made
out of a hole (empty site) plus a 
surrounding cloud of spin distortions that carry the actual quasiparticle
 spin. 
Depending on parameters the quasiparticle size
can involve dozens of sites, and thus it differs drastically from a
bare one-site hole. It has been shown that the use
of  quasiparticle
operators~\cite{quasi} provides more accurate information than 
other approximate
methods based 
on the bare degrees of freedom. 
Note that the distinction arises from the approximate
character of most calculations. If one were to solve the problems
exactly, the results would be independent of the
 actual initial formulation of the problem.

The goal of the present paper is twofold. Motivated by the discussion above,
first  a computational method is
discussed where the Lanczos method is applied on a reduced Hilbert space
to allow for the study of clusters larger than those that can be
handled exactly. To achieve this goal in an efficient way it is proposed
to change the basis in which the problem is formulated to accelerate the
convergence to ground state properties as the size of the reduced basis
grows (Sec.II). This basis can be improved systematically. The technique is
tested in hole undoped and doped models (Sec.III). 
The main advantage of this
method is that it provides dynamical information on intermediate size
clusters. The second goal of the paper is to apply this technique to the
study of the dynamical spin structure factor of doped 2-leg ladder
systems (Sec.IV). Clusters with $2 \times 16$ sites and 0 and
 2 holes were analyzed.
In agreement with previous studies~\cite{troyer}, it is here observed that the
excitation of lowest energy in the subspace of 2 holes and spin-1
corresponds to the bound state of the 2-holes pair (which already
exists in the
spin-0 sector) and a spin-triplet. Implications of this result for
neutron scattering experiments in both ladders and planes are 
briefly discussed (Sec.IV.B).

\section{Method}

As discussed in the Introduction, one 
of the purposes of this paper is to propose an improvement
on the method of diagonalization in reduced Hilbert
spaces~\cite{before}. This
method usually has the problem that the convergence to an accurate
result is slow as the reduced Hilbert space size grows~\cite{wenzel,jose}.
This complication may be caused by the improper selection of the
basis in which the problem is formulated which, thus far, has been
suggested mainly by the actual form of the Hamiltonian.
To improve on this approach let us consider as
 example the $t-J$ model on a 2-leg ladder. This
system is currently widely studied both
theoretically and experimentally due to the presence of a robust spin
gap in the spectrum and superconductivity upon doping~\cite{science}. In the
context of ladders it is easy
 to construct a good basis for 
the ground state. In fact
several studies have observed  that the ground
state of ladders in the realistic regime where the chain ($J$) and rung
($J_{\perp}$)
couplings are similar and at low hole-density
 is qualitatively related to the ground state in the 
large $J_{\perp}/J$ limit, which is formed by the direct product of
rung singlets~\cite{science}. This information suggests that the use of
a ``rung'' basis (made out of a singlet and a triplet per rung in the
hole undoped case) would 
be more suitable
than the standard so-called $S^z$-basis of spins up and down at each
site~\cite{subir}. The case of large $J_{\perp}/J$ makes this statement more clear: in the
$S^z$-basis, and working on an undoped $2 \times L$ ladder, 
$2^L$ states are still needed for the ground state. However, in the rung-basis the same
ground state is represented by just $one$ state which is the
configuration
with
singlets at every rung. As the ratio $J_{\perp}/J$ decreases
more states will be needed for the ground state to reach
a given accuracy in the energy or correlations, but
it is expected that their number will remain smaller than those needed
in the $S^z$-basis.

Although the basic idea of improving the basis in which a problem
is formulated has been used before in several contexts, such as in Physical
Chemistry, its application to problems of strongly correlated electrons
has been more limited. The DMRG technique mentioned before is an 
approach in the same spirit, improving the basis used in the problem at
hand, and it is
applied mainly to 
systems with open boundary conditions. The method described in the
present paper is a complement to such previous efforts, and provides an
alternative technique that allows for the calculation of dynamical
information and momentum dependent observables.

The method certainly goes beyond the mere change from the $S^z$-basis
to the rung-basis described in the previous paragraph. For instance, 
still in the context of ladders it
is natural to continue increasing the size of the clusters
which are considered exactly in the construction of the new basis. Next 
in line is
the plaquette basis built upon the exact solution of a $2 \times 2$ 
cluster~\cite{jorge} (see Fig.1a). 
Certainly the plaquette basis will
be even better than the site- and rung-basis since extra short distance
correlations are treated exactly in the problem.
In the plaquette basis the diagonal
energies $\epsilon_d$ range from $-2J$ to $J$ (for an undoped plaquette), 
and it is expected that the states will have a smaller contribution to the
ground state as their energies $\epsilon_d$ move away from $-2J$.
Numerically the
procedure can be continued using exactly solvable clusters of
increasing size.

\begin{figure}[htbp]
\centerline{
\hspace{3.5cm}
\psfig{figure=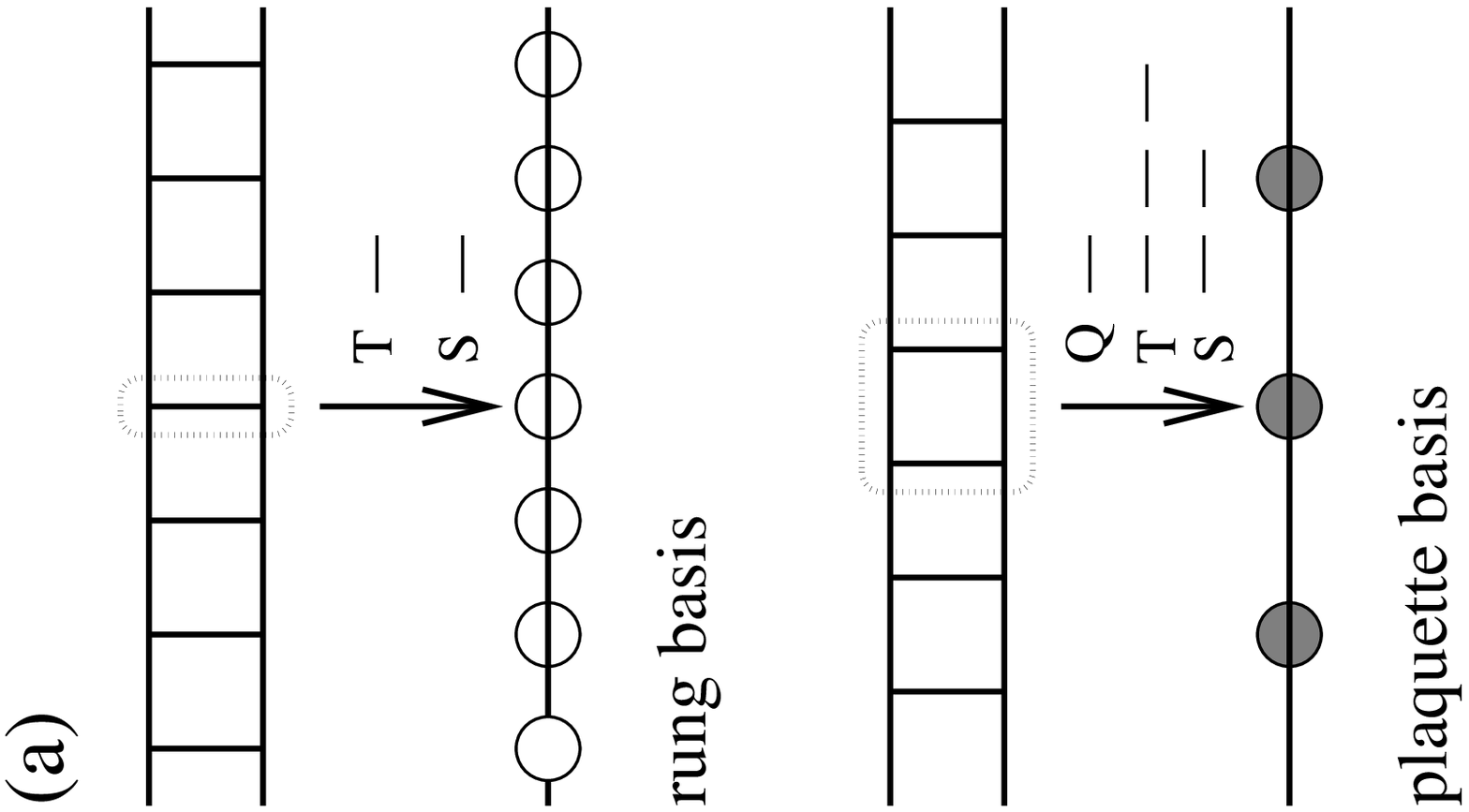,width=3.cm,height=7.0cm,angle=-90}
\psfig{figure=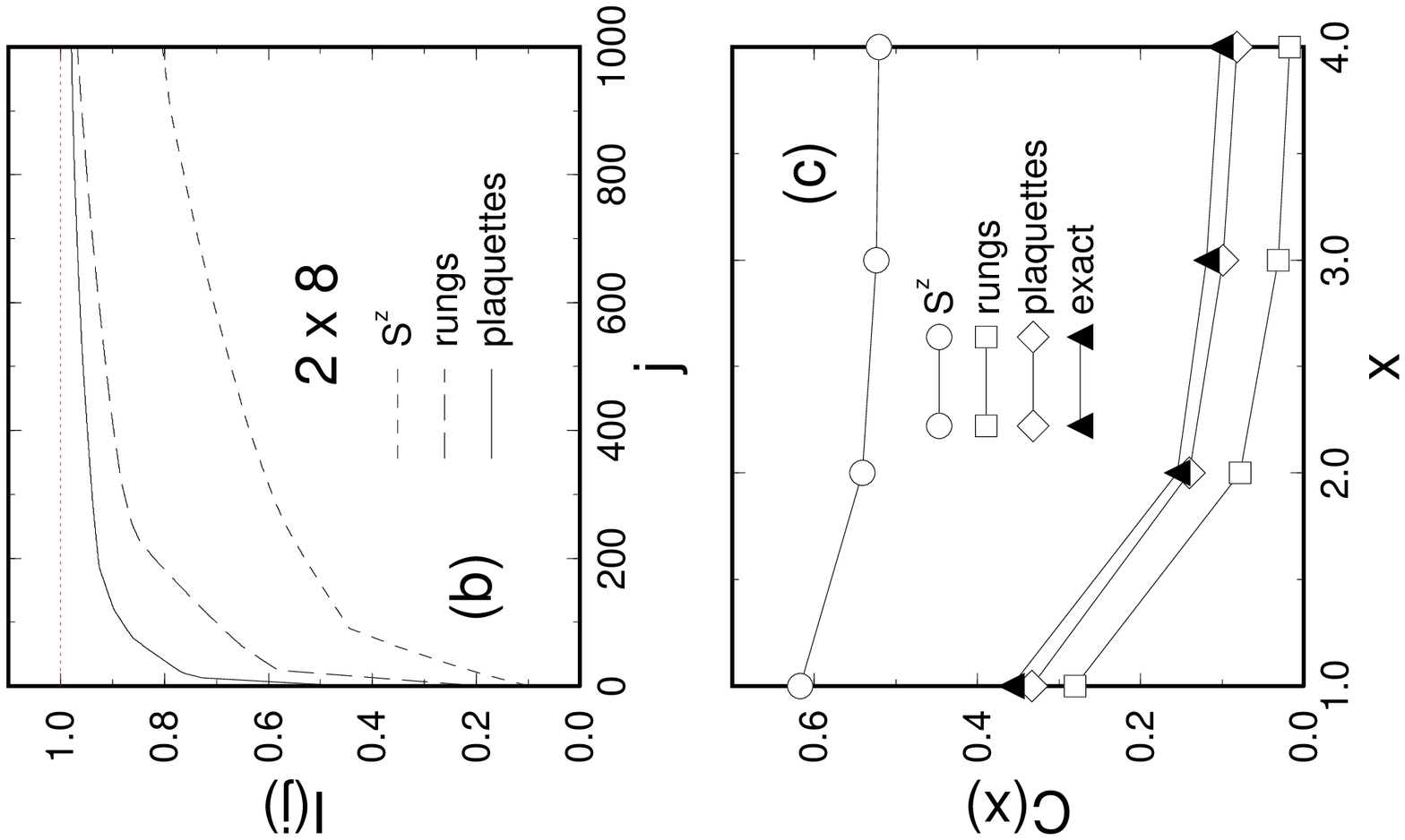,height=7.0cm,width=8.0cm,angle=-90}}
\vspace{0.5cm}
\caption{(a) Schematic representation of the change of basis to rungs and
plaquettes for an undoped system. 
Shown also schematically are the states for each basis with their spin
(S, T, and Q, denote a singlet, triplet, and quintuplet, respectively,
and the short segments next to them denote their number in the basis
under discussion, not their
energy);
(b) Integrated weight $I(j)$ (see text) vs  $j$, index 
that labels the states ordered sequentially from large to small
according to their weight in
the exact ground state of an undoped 
$2 \times 8$ cluster. Shown are results for 
three basis; (c) Staggered spin-spin correlation $C(x)$ for the same
 $2 \times 8$
cluster using the $2^8$ states which have the highest weight in the
exact ground state for the three basis indicated, compared with the
exact results obtained with the total 12,870 states.}
\label{fig1}
\end{figure}

The same reasoning can be applied to other spin models
such as those corresponding to dimerized chain systems where
the natural basis
is made out of singlets and triplets in the ``strong'' bonds. 
In 1D electronic and spin Hamiltonians, 
a good basis arises from solving exactly
$1 \times L$ small segments, as shown explicitly below. 
The same is true
whether one- or three-band Hubbard models are considered.
The physics of 2D systems can be approached
by the study of $N$-leg ladders with increasing value of $N$, as
recently proposed in DMRG calculations~\cite{doug4}. In this case
the block that should be solved exactly is the generalized $N$-site rung.
Thus, by no means the approach presented here is restricted only to
ladders, but it is general and independent of the cluster geometry and
model (results in 2D will be presented in future publications).
In general, the goal is to consider exactly at least part of the
 $short$ distance correlations by a suitable change of basis, and
then carry out some other numerical or analytical approximate 
technique to complete the
calculation for a given accuracy in observables.
By changing the basis
a better starting point for most 
many-body approximate techniques will be achieved~\cite{wilson1}.
Although it is unlikely that in 
quasi-1D systems the method will be more accurate than the DMRG
technique for equal-time correlations, nevertheless 
it can produce dynamical and momentum dependent correlations
and in this respect it is a complement to previous DMRG and Lanczos approaches.

\section{Tests of the technique}

\subsection{Undoped Systems}

Fig.1b illustrates the advantages of using basis that diagonalize
exactly a small cluster
over the $S^z$-basis for the case of a small undoped ladder. 
Diagonalizing the Heisenberg Hamiltonian for the $2 \times
8$ cluster in the three considered basis, the 12,870 states were 
sequentially ordered from large to small
according to their
weight in the normalized  exact ground state.
The integrated weight 
$I(j) = \sum_{i=1}^j  |c_i|^2$, where $c_i$ is the coefficient of the
$i$-th state in the ground state, provides information on the weight
distribution. A rapid convergence to 1 implies
that a small subset of the basis can carry an important
fraction of the total probability. 
In agreement with the previous discussion
the convergence is rapidly improved as the size of the cluster
exactly considered in the basis increases. For instance,
to reach 90\% of the total weight the plaquette basis needs about 150
states, while in the $S^z$-basis over 2,000 states are required
(results as encouraging as these ones are shown below also for
1D systems and doped ladders). Larger clusters beyond plaquettes would
improve even further this result.

\subsubsection{Equal-time Spin Correlations}

This basis-dependent redistribution of weight has consequences
for the calculation of expectation values when only a fraction
of the total Hilbert space is used.
As example consider
the staggered spin-spin correlations defined as
$C(x) = (1/N)
\sum_y (-1)^{x} 
\langle { {{\bf S}_{ y}}\cdot{{\bf S}_{y+x}} } \rangle$ (standard
notation) which are of special
importance since their long distance behavior reflects on the
gapless vs gapped character of the energy spectrum. 
It would be desirable that a reduced basis  provides a
qualitatively correct $C(x)$ at large distances.
Considering only the 256 states with the largest coefficients
for the $2 \times 8$ ladder, Fig.1c shows that the
$S^z$-basis gives a staggered spin correlation  too  large 
compared with the exact result. This problem  occurs 
because the states with the 
largest ground state weight in this basis
are small modifications around
 the pure N\'eel state which has staggered order. 
On the other hand, using the
basis that diagonalize blocks with 2 and 4 sites 
considerably better results are
achieved with the same computational effort 
since the dominant state in the rung-basis 
(singlets in all rungs) already has a robust spin-gap
as the exact ground state of the 2-leg ladder.

\begin{figure}[htbp]
\centerline{\psfig{figure=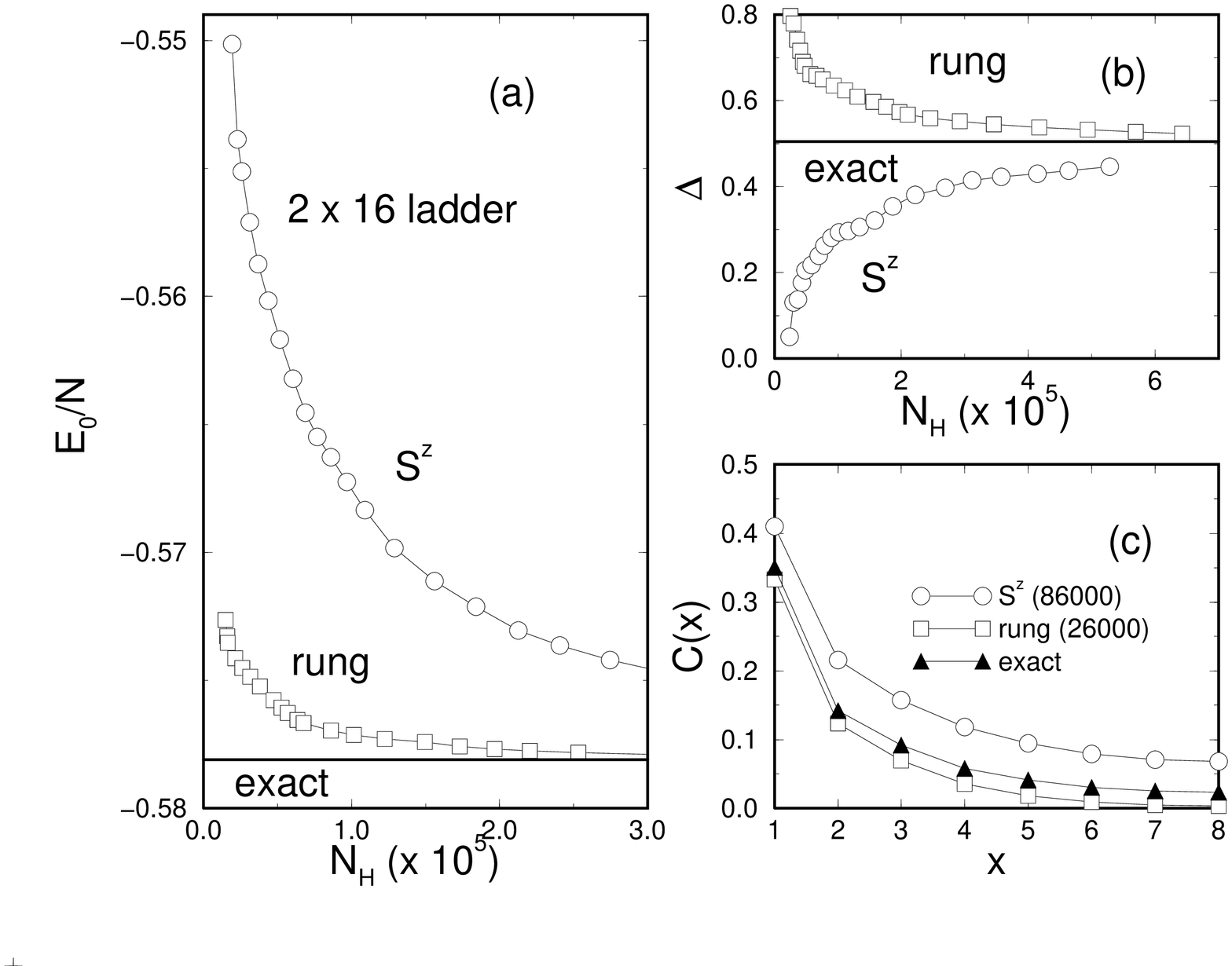,width=8.cm,height=7.0cm,angle=-90}}
\vspace{0.5cm}
\caption{(a) Ground state energy per site $E_0/N$ (in units of $J$)
 vs size of the reduced
Hilbert space $N_H$ for the Heisenberg model on
a $2 \times 16$ ladder using the $S^z$- and rung-basis. The exact
(Lanczos) result  is $-0.5781032$;
(b) Spin-gap vs $N_H$ on an undoped
$2 \times 16$ ladder using the $S^z$- and rung-basis. The exact result
$0.505460384$ is also shown. Each
point represents the difference between the energies of the lowest
energy state in the subspaces of spin-1 and -0
for a given size of the Hilbert space
common to both 
subspaces; (c) Staggered spin-spin correlation $C(x)$
for the $2 \times 16$ cluster. 
Results in the
$S^z$-basis (rung-basis) were obtained using $\sim 86,000$ ($\sim 26,000$)
states. The exact results were obtained with the Lanczos method.}
\label{fig2}
\end{figure}

In Fig.2a the evolution of the ground state energy as the dimension of
the reduced Hilbert space grows is shown for the $S^z$- 
and the rung-basis using now an undoped $2 \times 16$ cluster.
A considerably faster convergence is achieved with the latter producing
4 significant figures in the energy with only 300,000 states
($\sim 1.6\%$ of the total space).
The details of the expansion procedure are independent of the basis
used, and since they are described in previous literature they will not
be repeated here~\cite{jose}.
Similar rapid convergence is achieved if the energy of the first excited
state of spin 1 and momentum $(\pi,\pi)$ is investigated
(Fig.2b).
Regarding $C(x)$ the behavior on the undoped $2 \times 16$
cluster (Fig.2c) resembles results on smaller systems:
better correlations are obtained with the rung-basis even if less states
are used than in the $S^z$-basis.

The algorithm described in this paper is used for clusters with
periodic boundary conditions, and thus it is possible to carry out the
basis expansion procedure in a basis of momentum eigenstates. This allows us
to gather information about, e.g., excited states of spin 1 with
different momenta which is important to compare results against those
obtained with inelastic
neutron scattering techniques. Results for the Heisenberg model are
given in Fig.3a. Once again using a small fraction of the total space
and the rung-basis
accurate energies are obtained at all momenta for this cluster. The
results are in excellent agreement with previous predictions by Barnes
and Riera~\cite{br}.

\begin{figure}[htbp]
\centerline{\psfig{figure=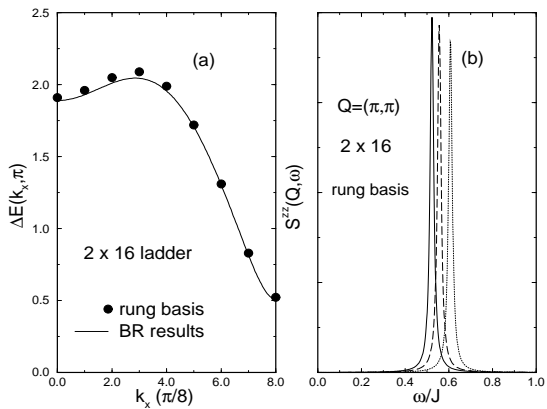,width=7.5cm,height=6.0cm,angle=-90}}
\caption{(a) Energy gap between the state of lowest energy in the subspace 
of spin-1 and momentum-${\bf q}$ and the ground state for the Heisenberg model, using 
$\sim 300,000$ rung-basis states in each subspace. The rung momentum is fixed to $\pi$.
Results are compared against the calculations of Barnes and Riera (BR)
(b) Dynamical spin structure factor $S^{zz}({\bf Q}, \omega)$ with ${\bf Q} =  (\pi,\pi)$. 
Dotted, dashed, and solid lines indicate results using $120,000$, $253,000$, and $410,000$ states in the reduced Hilbert 
space of spin-1 and momentum ${\bf Q}$ obtained after the application of
the $S^z_{\bf Q}$ operator over a reduced ground state with about $50,000$ rung-basis states. The latter was obtained from
an approximate overall ground state (spin-0) of about $600,000$ rung-basis states.}
\label{fig3}
\end{figure}

\subsubsection{Dynamical Spin Structure Factor}

An interesting advantage of the method proposed here is that
having a good
 approximation to the ground state 
expressed in a simple enough  basis allow us to
obtain $dynamical$ information without major complications.
In other techniques
the ground state is either lost in the iterative processes,
or it is expressed in a cumbersome basis for
the application 
of the operator being investigated
(spin, charge, current) over such ground state.
 As example,
 here the dynamical spin structure factor
$S^{zz}({\bf q}, \omega)$ was calculated. The actual procedure is
simple: the operator ${\hat O} = S^z_{\bf q}$ (standard notation)
 is applied to the ground
state in the reduced basis denoted by
$| \phi_0 \rangle_{T}$.
If all states of the subspace of spin-1 and momentum ${\bf q}$ generated
by the operation ${\hat O} | \phi_0 \rangle_{T}$ were kept in the
process, typically one would excess the memory capabilities of
present day workstations if the truncated ground state has about
$3 \times 10^6$ states. Then, it is convenient to work with just a
fraction of $| \phi_0 \rangle_{T}$, say keeping about 10\% of the
states. In this way the subspace of spin-1 under investigation
typically has a similar size as the original reduced basis ground state,
namely approximately $3 \times 10^6$ rung-basis states.
The state ${\hat O} | \phi_0 \rangle_{T}$
constructed by this procedure
 is now used as the starting configuration for a standard
continued fraction expansion generation of the dynamical response
associated to ${\hat O}$~\cite{review}. 
As a test results are shown in Fig.3b for an undoped $2 \times
16$ cluster and ${\bf q } = (\pi,\pi)$. As observed in the figure,
for the case of the undoped 2-leg ladder
$S^{zz}({\bf q},\omega)$ is dominated by just one peak with small weight 
at higher energies. The convergence as the reduced basis set
grows is fast, and the results are quantitatively accurate even with 
only $\sim 1\%$ of the total space in the ground state. 
Results at other momenta and for
doped ladders behave similarly and they
will be analyzed later in Section IV.

\subsubsection{Results for Intermediate Size Clusters}

Fig.4 contains $C(x)$ for the Heisenberg model on a $2 \times 20$
 cluster, which cannot be studied exactly. Using up to 
$\sim 1,600,000$ states (just $0.04 \%$ of the total space)
the results for $C(x)$ are in good
agreement with world-line Monte Carlo simulations at low temperatures. The ground state
energy is obtained with three significant figures using this basis set.
The existence of a
short antiferromagnetic correlation length $\xi_{AF}$~\cite{science,wns94}
is clear from Fig.4, and
in this respect 
the basis expansion method in the rung-basis
 has captured properly the qualitative aspects
of the ladder ground state, which are dominated by short-range
antiferromagnetic fluctuations and rung singlet formation. 
Similar calculations in the $S^z$-basis keeping the same number of
states incorrectly suggest that
$\xi_{AF}$ is very large.
In addition, note that
there are undoped ladder compounds, such as ${\rm Cu_2 (C_5 H_{12} N_2 )_2 Cl_4
}$, that are highly anisotropic with ${ J_{\perp}/J =
5.5}$~\cite{aniso}. 
In this regime the rung-basis is
particularly useful: combining information from
exactly solvable clusters
and  Lanczos in a reduced rung-basis set  for larger systems,
the spin-gap for 
${ J_{\perp}/J = 5.5}$
was accurately found to be $4.598105.$

\begin{figure}[htbp]
\centerline{\psfig{figure=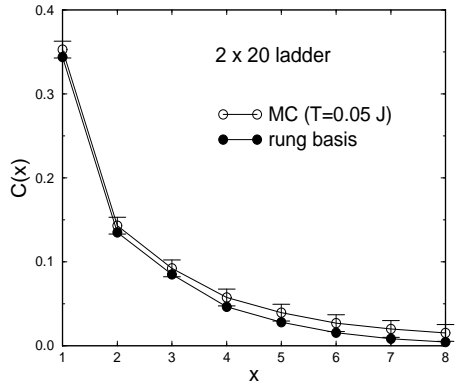,width=8.cm,height=7.0cm,angle=-90}}
\caption{Spin correlation $C(x)$ vs $x$ for a $2 \times 20$ Heisenberg
cluster obtained keeping 
$1,600,000$ states in the reduced basis ($0.04 \%$ of the total Hilbert
space), compared with world-line Monte Carlo results for the same
cluster at temperature $T=0.05J$.}
\label{fig4}
\end{figure}

\subsection{Doped systems}

The method proposed here certainly applies to hole doped (fermionic) systems,
and to models in geometries different from ladders. To illustrate these
cases in Fig.5a-b results are presented for (i) a 1D $t-J$ chain  with 2 holes
using as basis the states that diagonalize exactly
blocks of 2 and 4 sites, 
and (ii) an anisotropic ladder with 1 hole using the rung
basis (and with an anisotropy corresponding to the region where
pairing correlations are maximized upon further doping~\cite{noack}). 
The advantage of using these new basis is clear from the figure.
For the anisotropic ladder $80\%$ of the weight is obtained with 15
states in the rung-basis, compared with 350 in the $S^z$-basis.
In addition, Fig.5c shows results for the $t-J$ model with 2 holes on 
a $2 \times 16$ cluster that cannot be studied exactly. Using $\sim
500,000$ rung-basis states a ground-state snapshot
is provided in Fig.5c where one hole is projected to be at
 an arbitrary site
and the probability of finding the second hole at some other site
is presented. The results illustrate the formation of two-hole bound
states and they are in excellent agreement with those previously 
reported using other techniques~\cite{doug4,riera2}.
Then, it is concluded that the method discussed in this paper can handle
properly systems with holes, in addition to undoped spin models.

\begin{figure}[htbp]
\centerline{\psfig{figure=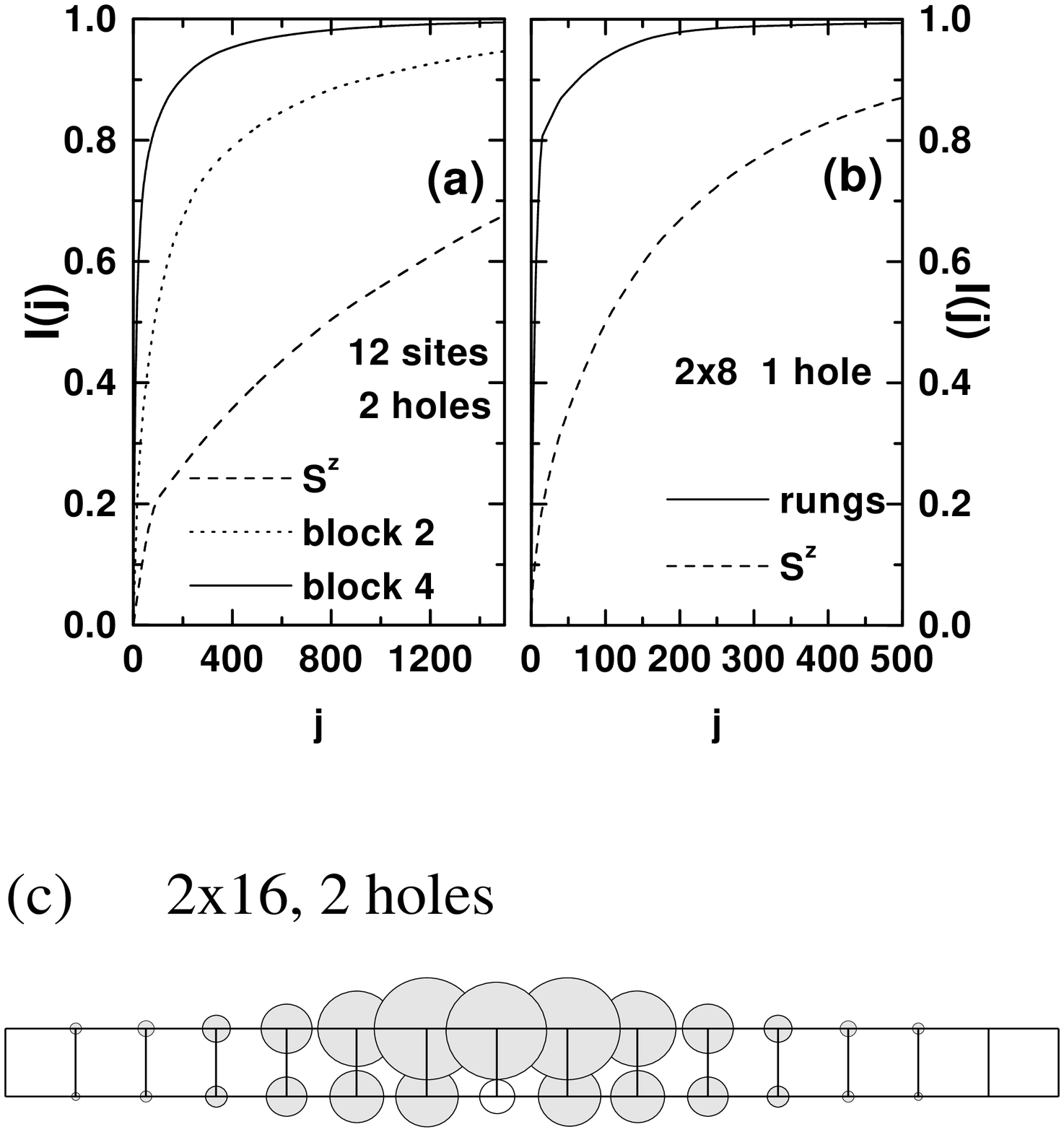,width=7.cm,height=7.0cm}}
\vspace{1.cm}
\caption{(a) $I(j)$ vs $j$ (see Fig.1b) using the exact ground state of
the $t-J$ model on a 12-site chain with 2 holes at 
$J/t= 0.4$. The basis are the
standard $S^z$, and the 2-block and 4-block basis where the
exact solutions for clusters of 2
and 4 sites are used; 
(b) $I(j)$ vs $j$ using the exact ground state of
the $t-J$ model on a $2 \times 8$ anisotropic ladder with
$J/t =0.4$, $t_\perp /t = 1.5$, $J_\perp /t = 0.9$  and 1 hole. Results for the
$S^z$- and rung-basis are shown; 
(c) Distribution of holes for the
$t-J$ model on a
$2 \times 16$ cluster with 2 holes at $J/t = 0.4$ using 
$\sim 300,000$ rung-basis states. The area of the gray circles at a given site
are proportional to
the probability of finding a hole at that site, once the other hole is
fixed at an arbitrary position (open circle)}
\label{fig5}
\end{figure}

\section{Dynamical Spin Structure Factor in Doped Ladders}

\subsection{Results and Interpretation}

As an application of the technique presented in the previous Sections,
here an analysis of $S({\bf q}, \omega)$ will be reported for lightly
doped 2-leg ladders. The study is performed on clusters of size $2
\times 16$ which cannot be studied exactly. Previous 
work in this
context
discussed an interesting property of the subspace of spin-1 in doped
ladders~\cite{troyer}. The overall ground state in this subspace corresponds either to two
unbound hole quasiparticles (each carrying a spin-1/2) or to a $bound$
state of a hole pair 
(similar to the pair that appears in the ground state of the
2-holes spin-0
subspace~\cite{science}) and
a spin-triplet. This result should be contrasted against the
ground state of the spin-1 sector in undoped ladders which in the limit
of a large rung exchange $J_{\perp}$ simply
consists of one spin-triplet at an arbitrary rung and spin-singlets in all the
other rungs. Such a result  suggests that in doped ladders with 2 holes the
state with a tight hole-pair separated in space from the rung-triplet
would contribute substantially to $S({\bf q},\omega)$. This is indeed
true and the spin dynamical structure factor is dominated by such
a configuration. However, the state where the pair and the spin-triplet
are bound has a lower energy and it is expected to
generate a low-intensity branch
of excitations in the spectrum of lightly doped ladders, 
below the high-intensity branch that
evolves smoothly from the undoped limit~\cite{troyer}.

In Fig.6, $S({\bf q},\omega)$ is presented on a $2 \times 16$ cluster
with no holes, and using
the technique described in this paper. In this study efforts are
concentrated on the $q_y=\pi$ branch which is the most interesting for the
physics of the problem. The spin-0 ground state has here $\sim 10^6$
states, of which the dominant $\sim 150,000$ where used in the dynamics.
The subspace obtained by the application of the $S^z_{\bf q}$ operator
over this
state also has approximately $10^6$ rung-basis states.
A clear peak can be observed in the figure at
all momenta, with
a pole energy position already provided in Fig.3a. The largest weight is
obtained at momentum $(\pi,\pi)$.

\begin{figure}[htbp]
\centerline{\psfig{figure=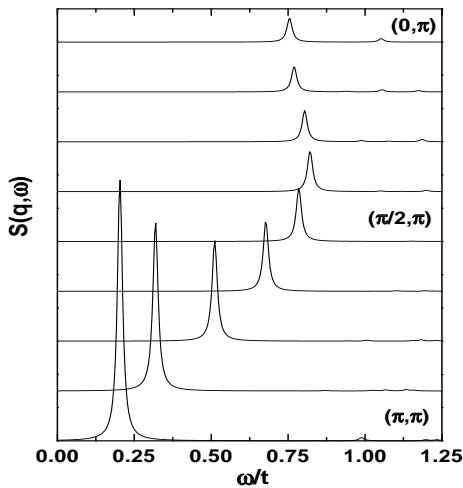,width=7.5cm,height=8.0cm}}
\caption{$S({\bf q},\omega)$ for the $q_y = \pi$ branch on a $2 \times 16$
cluster with no holes. 
The technique used is the method proposed here
in Sec.II. The number of states in the reduced basis is explained in the
text.  The $\delta$-functions have been given a width $0.01t$.}
\label{fig6}
\end{figure}

Fig.7 contains a similar result but now
in the sector of 2 holes using $\sim 3 \times 10^6$ rung-basis states
for the spin-0 ground state, of which $\sim 150,000$ are kept for the 
dynamics. The subspace of 2 holes, spin-1 and momentum ${\bf q}$
typically has also $\sim 3 \times 10^6$ states.
The spectrum has two main branches: (I)
is the branch with the highest intensity and it is in good
correspondence energy- and shape-wise with the result in the undoped
limit. As explained before, these states are expected to correspond to a
tight pair of holes plus a spin excitation, separated spatially from the
first (i.e. without forming a bound state); (II) corresponds to the
actual ground state of the subspace of spin-1 at all the momenta
investigated here. This
state is expected to 
be a bound state between a hole-pair and a spin excitation.
Note that symmetry considerations~\cite{troyer} explain why 
branch (II) has zero
weight at momentum $(\pi,\pi)$. The reason is that with 
two holes the actual 
lowest-energy state of spin-1 is exactly orthogonal to the state 
obtained from applying $S^z_{\bf Q}$ to the spin-0 2-hole ground state. These states transform
differently under reflections along the ladder direction. Such an effect
is expected to disappear for a larger number of holes. In spite of this
orthogonality problem, the position of the bound state pair-triplet at
momentum $(\pi,\pi)$ can be obtained using other techniques. For
instance, using the DMRG method on a $2 \times 16$ cluster with
periodic boundary conditions, 2 holes, total $z$-projected spin one, 
and
$J/t=0.4$ it was found that the spin-gap is $\sim 0.26t$ ($m=200$,
truncation error $5 \times 10^{-5}$). Information
about this gap can actually also be found in the continued fraction
expansion procedure to obtain Fig.7
since a pole with negligible weight nevertheless
appears in the process. Its position is very similar to the result found
with DMRG.

\begin{figure}[htbp]
\centerline{\psfig{figure=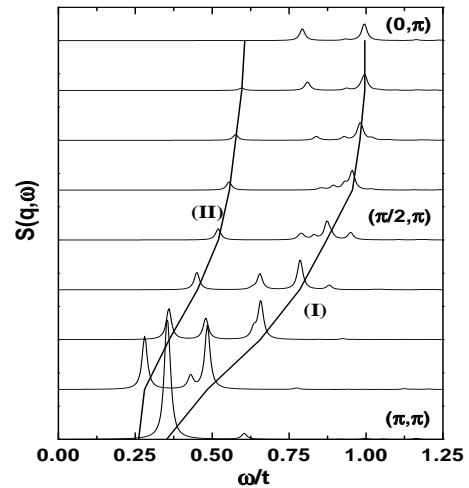,width=7.5cm,height=8.0cm}}
\vspace{0.5cm}
\caption{Same as Fig.6 but for the case of a $2 \times 16$ ladder with two
holes. The physical
meaning of branches (I) and (II), as well as the number of states used
in the study, are explained in the text.}
\label{fig7}
\end{figure}

To verify in more detail the physical interpretation of branch (II), 
the ground state of an $2
\times 8$ cluster was obtained exactly using
the Lanczos algorithm working in the rung-basis. 
The dominant configuration in the limit where
$J_{\perp}$ is the largest scale is shown in
Fig.8a (using as example $J_{\perp}/J = 10$, $J/t=0.4$, and
hopping amplitudes $t=1$ along both rungs and legs). 
This state has
two holes in the same rung, and a rung-triplet right next to the pair 
(of course the state with the triplet on the other
side of the pair carries the same weight). 
The next dominant state is shown
in Fig.8b and it corresponds to having nearest-neighbor rungs with one
hole and a spin up each. The overall $q_y = \pi$ momentum is obtained by using one hole with rung-momentum 0 and the other with $\pi$. 
The combination of the states Fig.8a and 8b (and
the other two obtained by reflections along the legs) carry $74\%$ of
the weight of the ground state in this large $J_{\perp}$ example. 
The fact that both the pair of holes and
the spin-1 are located spatially close to each other indicates that a
bound state pair-triplet has been formed, as anticipated.

\begin{figure}[htbp]
\centerline{\psfig{figure=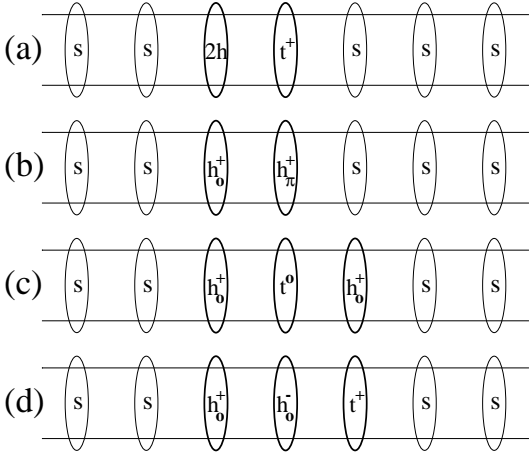,width=7.cm,height=6.0cm,angle=-90}}
\vspace{0.5cm}
\caption{(a) Dominant configuration in the rung-basis corresponding to a
$2 \times 8$ ladder with $J_{\perp}/J = 10$, $J/t=0.4$, and
hopping amplitudes $t=1$ along both rungs and legs;
(b) Next dominant
configuration for the same parameters as in (a); (c) Dominant configuration
for the isotropic case $J_{\perp}/J = 1.0$, with all other parameters as
in (a); (d) Next dominant configuration for the parameters of (c).  Both
in (c) and (d) the results are the same for $2 \times 8$ and $2 \times
16$ clusters. The convention
followed here is the following: $s$ denotes a spin
singlet along the rung, $t^+$ is a rung-triplet with $z$-projection +1,
$t^0$ is a rung-triplet with $z$-projection 0, $2h$ is the state of two
holes in the same rung, and $h^+_0$ and $h^+_{\pi}$
are the rung-states of one hole with spin-up and momentum $0$ and $\pi$,
respectively.}
\label{fig8}
\end{figure}

Fig.8c shows the dominant configuration for the more realistic case of
$J_{\perp} = J = 0.4$ and $t=1$. 
The holes forming the pair are 
now located at a distance of two lattice
spacings along the legs, with a spin-triplet in between. 
In the next dominant configuration (Fig.8d)
the holes are separated by one lattice spacing, with the triplet on the side.
These states still represent the bound state
pair-triplet observed at large $J_{\perp}$ and predicted in
Ref.~\cite{troyer}. The main differences between Figs.8a-b and c-d 
simply originate in the fact that the bound state is more extended in
the rung-leg isotropic limit than in the $J_{\perp} \gg J$ case.
It is interesting to observe that these configurations are also
dominant for larger clusters. Actually the expansion procedure described in
this paper was applied to the $2 \times 16$ cluster, in the spin-1
subspace, with momentum $(\pi,\pi)$, keeping about $10^6$ states. The
dominant rung-basis states were found to be
 the $same$ as shown in Figs.8c and d. Note also the clear advantage of
using the rung-basis in this visualization of dominant configurations:
the state of Fig.8a expressed on a $2 \times 16$ cluster and using the
$S^z$-basis would require $2^{14}= 16,384$ states.

\subsection{Implications for Neutron Scattering Experiments}

\subsubsection{Ladders}

The results of the previous subsection have implications for neutron scattering
experiments on ladder compounds. 
Although the study of ladder materials
 with neutron scattering techniques has been restricted thus far to
the undoped limit~\cite{ecle}, experiments for doped ladders 
${\rm Sr_{14-x} Ca_x Cu_{24} O_{41} }$ are being 
planned~\cite{ecle2}.
In this context, experimentalists 
should observe two branches in their spectra, namely (I) and (II)
of Fig.7. Branch (II) has an intensity that grows with the number of
holes, and thus naively it is expected to be weak. However, note 
that in Fig.7 and considering momentum ${\bf q} = (7 \pi/8, \pi)$, the weights
of both branches (I) and (II) are $similar$ in intensity, even in a situation where
 the nominal density is as small as
$x = 2/32 = 0.0625$. Thus, the experimental search for branch (II) may not
be difficult, and its observation would provide  support to
calculations that predict the formation of tight pairs in 2-leg 
ladders~\cite{science}.

\subsubsection{Planes}

Note that excitations such as those corresponding to branch (II) could
appear also in the two-dimensional high-Tc cuprates. It is well-known that
these materials present a pseudogap behavior in the underdoped regime.
Some theories explain this feature as caused by 
magnetic effects~\cite{magnetic}, while
others attribute its origin to preformed hole pairs~\cite{preformed}. 
A mixture
of these two proposals was recently introduced using results on ladders as 
a guidance~\cite{martins}.
In this context it was suggested  that 
 the short-range antiferromagnetic (AF) order
of lightly doped cuprates
may lead to long-lived
hole-pairs in the normal state. These pair may cause several
anomalies in transport, neutron, and photoemission experiments. In this
framework the pseudogap is correlated with  the existence of a finite
AF correlation length, which is necessary to form the pairs.

In Ref.\cite{martins} the
spectral function of doped ladders for up to $2 \times 20$ sites
clusters was presented. A gap at $(\pi,0)$ was
clearly observed due to the existence of 
tight pairs on ladders. Analyzing the same $t-J$ ladder model and 
hole density, both in 
Ref.\cite{troyer} and here
it has been observed that having a state with
tight hole pairs have important consequences not only for photoemission
experiments but also for neutron-scattering experiments
since in this context a new branch in the spectra should be observed 
(tight spin-triplet hole-pair).
Note that the effect discussed here is not caused by any bilayer structure
as some theories propose~\cite{theories1},
but it is expected to appear in isolated copper-oxide planes 
as well. 
Also there is no need to fine
tune parameters in the $t-J$ model to observe the states
describe in this section, which exist in a wide region of parameter space.

These novel excitations may have already been observed in
 the broad normal state peaks of underdoped YBCO reported 
recently~\cite{exper}, related to the famous resonances
in the superconducting state of optimally doped YBCO~\cite{optimal},
but certainly considerable 
more work is needed to relate the present ideas with experiments.
It also remains to be investigated what is the relation
between the bound pair-triplet excitation branch (II),
and other theories for the sharp
peaks in neutron scattering that appear in cuprates,
such as the
collective modes in the particle-particle channel~\cite{zhang}
and the spin-wave excitation explanation~\cite{pines}.

\section{Summary}

Summarizing, here it was proposed that the accuracy of some
approximate techniques for the study of many-body problems, such as the
 Lanczos approach in a reduced basis, can be improved
if  a change of basis is performed that incorporates exactly short
distance interactions in the problem. 
Numerical methods using reduced
Hilbert spaces
converge faster in the new basis than in those naively suggested by
the degrees of freedom explicit in the Hamiltonian. 
The method described here is
certainly  in
the same spirit as previous techniques that also improve systematically the
basis used in a given problem, such as the DMRG method. The advantage of
the approach discussed here is that it allows for the calculation of
dynamical information on clusters larger than those that can be solved exactly.
Although the technique was 
tested only
on undoped and doped ladder and chains 
systems, the approach can be used for a wide
variety of problems independent of their dimension and geometry, and for
both doped and undoped systems.
Analytic techniques, such as perturbation theory, can 
also improve their convergence radius using a better basis~\cite{wilson2}.

In addition, in the present paper the dynamical spin structure factor
$S({\bf q}, \omega)$ was calculated for lightly doped ladders. In
agreement with previous literature~\cite{troyer}, a low-intensity branch
was observed away from half-filling. This branch appears at an energy
smaller than the high-intensity branch that smoothly evolves from the
undoped system. This new branch is characteristic of systems with tight
hole pair bound states, and it should be detectable in neutron
scattering experiments for ladders. If hole pairs (or
long-lived pairs)
exist in the normal state of the underdoped two-dimensional cuprates,
such an excitation could be detectable also in this context.

\section{acknowledgments}

E. D. thanks the financial support of the NSF grant DMR-9520776. G. M.
and A. M. thank
the Conselho Nacional de Desenvolvimento Cient\'\i fico
e Tecnol\'ogico (CNPq-Brazil) for support, as well as 
the NHMFL
In-House Research Program, supported under grant DMR-9527035.
C. G. thanks CONICET, Argentina, for support.

\end{document}